\documentclass[a4]{article}
\usepackage{authblk}
\usepackage{graphicx}
\usepackage{amsmath}
\usepackage{amsfonts}
\usepackage{amssymb}
\begin{document}

\title{Trapping effects on the vibration-inversion-rotation motions of an ammonia 
molecule encapsulated in C$_60$ fullerene molecule.}
\author{Azzedine Lakhlifi\footnote{
Institut Universitaire de Technologie, Belfort-Montb\'{e}liard. 
E-mail : azzedine.lakhlifi@obs-besancon.fr} \\
Institut UTINAM-UMR CNRS 6213  Universit\'{e} de Franche-Comt\'{e} \\
Observatoire de Besan{\c c}on \\ 41 bis avenue de l'Observatoire - BP 1615 -
25010 Besan{\c c}on Cedex, France \\
and Pierre R. Dahoo \\
CNRS-INSU, LATMOS-IPSL - Universit\'{e} de Versailles \\
11 Bd d'Alembert - 78820 Guyancourt, France}
 
\maketitle

\begin{abstract}
The infrared bar-spectrum of a single ammonia molecule encapsulated in nano-cage 
C$_{60}$ fullerene molecule is modelled using the site inclusion model successfully 
applied to analyze spectra of CO$_2$ isotopologues isolated in rare gas matrix. 
Calculations show that NH$_3$ can rotate freely on a sphere of 
radius 0.184 $\text{\AA}$ around the site centre of the nano-cage and spin 
freely about its C$_3$ symmetry axis. 
In the static field inside the cage degenerate 
$\nu_3$ and $\nu_4$ vibrational modes are blue shifted and 
split. When dynamic coupling with translational motion is considered, the 
spectral signature of the $\nu_2$ mode is modified with a higher hindering 
barrier (2451 cm$^{-1}$), an effective reduced mass (6.569 g.mol$^{-1}$) and 
a longer tunneling time (55594 ps) for the fundamental 
level compared to gas-phase values (2047 cm$^{-1}$), (2.563 g.mol$^{-1}$) 
and (20.85 ps). As a result this mode is red shifted. 
Moreover, simulation shows that the changes 
in the bar-spectrum of the latter mode can be used to probe the temperature 
of the surrounding media in which fullerene is observed. 
\end{abstract}

\section{Introduction}

Quite recently, "buckyballs" have been observed for the first time in space by 
Astronomers of University of Western Ontario \cite{nasajully10,jcjbsepsem2010} 
using NASA's Spitzer Space Telescope. They identified spectral signatures 
of fullerenes in a cloud of cosmic dust surrounding a distant star 6500 light 
years away. The existence of fullerenes has thus been 
confirmed as anticipated after the first observation in laboratory by 
Kroto {\it et al.} at Rice University \cite{hwkjrhscobrfcres85}. One may therefore 
expect observation of molecular species trapped in fullerenes in interstellar medium. 
In this work we investigate on the changes that are induced in the spectral signatures 
of species trapped in fullerene in order to determine how these changes can be used 
to characterize the environment where fullerenes are observed. From inert matrix 
isolation spectroscopy, one knows that infrared (IR) spectra of molecular 
species trapped in a site are simplified in that, as rotational structures 
are absent in the spectra, a vibrational transition 
is observed as one peak blue or red shifted or sometimes two or more in case of 
splitting of degenerate levels \cite{cgal89_1,prdibvrjlthcallam99,prdalhcjmc2006}.

Trapping of atoms or ions and small diatomic molecules in fullerene have been studied 
both experimentally and theoretically 
\cite{dsbrdjjrsmsdvcsy93,hstwdkbjh93,mshajvrjcrjp93,hajvrjcmsrjp94,
mshajvrjcsmmlgdeg94,rsrjcms97,rjcakms2003,jc91,ciwmawlp93,jhrjbjmgl96_1,jhrjbjmgl96_2,
ehtoavdapesw96,rjc2008,mxfszbrlnjt2008_1,mxfszbrlnjt2008_2,zsfulasn2005}. 
However, except for CO \cite{jhrjbjmgl96_2,ehtoavdapesw96,rjc2008} 
and H$_2$ \cite{mxfszbrlnjt2008_1,mxfszbrlnjt2008_2} molecules, only a few 
of the theoretical work has been done in the infrared region. In the case 
of NH$_3$ trapped in fullerene, calculations have been performed 
from {\it ab-initio} methods to study the trapping 
mechanism \cite{zsfulasn2005,selt2003,mdgmmog2009} but no experimental 
work has been published. Erko\c{c} and T\"{u}rker \cite{selt2003} have 
reported that up to six NH$_3$ molecules can be trapped 
in C$_{60}$, on one hand and Ganji {\it et al.} \cite{mdgmmog2009} suggest 
that only one NH$_3$ molecule can form a stable 
complex NH$_3$@C$_{60}$ on the other hand. While Slanina {\it et al.} 
\cite{zsfulasn2005} have reported a binding energy of -1830 cm$^{-1}$ with 
the ammonia molecule oriented towards a pair of parallel pentagons.

The aim of this paper is to determine how fullerene modifies the spectral signature 
of ammonia molecule, in particular in the region of the vibration-inversion 
mode (namely umbrella mode). 

A theoretical study of an ammonia molecule trapped in fullerene C$_{60}$ is thus 
performed in order to simulate its infrared spectrum in the vibration-inversion 
frequency region.
The choice of NH$_3$ is motivated by the fact that it contains 
a nitrogen atom. CO$_2$ could otherwise have been chosen for the simulation.

To simulate the infrared spectra of NH$_3$ in fullerene, we use the site 
inclusion model successfully applied to analyze spectra of CO$_2$ isotopologues 
\cite{prdalhcjmc2006} isolated in rare gas matrix. When the molecule is 
trapped in the nano-cage of fullerene, the interaction with carbon atoms in the 
short range scale modifies the translational and rotational degrees of freedom 
because the molecule is no longer free to move. As a result, the umbrella mode 
which is characterized by the inversion doubling through tunneling of the 
nitrogen atom across the plane defined by the three H atoms is perturbed 
as discussed below. 

To calculate the bar-spectra of NH$_3$ trapped in C$_{60}$ nano-cage a 
renormalization procedure is applied on the total Hamiltonian of the system. The 
dynamical coupling between the molecular degrees of freedom and those of C$_{60}$ 
nano-cage is then small enough, at least to a first approximation, to justify 
an \textit{initial chaos hypothesis}. 
As a result, the density matrix operator of the system can be factorized into a 
product consisting of two terms, one related to the NH$_3$ renormalized optical 
states and the other to the C$_{60}$ bath states spanned by its vibrational modes. 
The dynamical line-shifts and line-widths of the bar-spectra can be disregarded and 
C$_{60}$ nano-cage considered as rigid. 

The interaction potential model used in our calculations is 
presented in section 2 and the renormalization method applied to separate 
the different motions from each other described. Results of the calculation 
to determine the frequency shifts of the vibrational modes and the orientational 
level schemes of the molecule are then given. Sections 3 and 4 are devoted to 
a presentation of the formalism used for the construction of the infrared 
spectra of NH$_3$ in fullerene and to a discussion of the infrared bar-spectrum 
in the frequency region of the $\nu_2$ mode.

\section{The interaction potential energy}

\subsection{Potential energy model}

The interaction potential energy $V_{\text{MC}}$ between the trapped ammonia molecule NH$_3$ 
and the rigid fullerene molecule C$_{60}$ is modelled as the sum of 12-6 Lennard-Jones (LJ) 
pairwise atom-atom potentials characterizing the repulsion-dispersion contributions and the  
induction part due to the interaction between the permanent electric multipole moments of the 
molecule and their images created at the positions of the polarized fullerene carbon atoms. 
It can be expressed as 

\begin{eqnarray}
V_{\text{MC}} = \sum\limits_{j=1}^{60}\sum\limits_{i=1}^44\epsilon_{ij}\left\{ \left(
\frac{\sigma _{ij}}{\left| \mathbf{r}_{ij}\right| }\right) ^{12} 
-\left( \frac{\sigma _{ij}}{\left| \mathbf{r}_{ij}\right| }\right) ^6 \right\} %
-\frac 12\sum\limits_{j=1}^{60} \textbf{E}_{\text{M}}^j :\mathbf{\alpha }^j :\textbf{E}_{\text{M}}^j, 
\label{eq1}
\end{eqnarray}

\noindent where \textit{i} and \textit{j} denote the \textit{i}th atom of the ammonia 
molecule and the \textit{j}th carbon atom of the fullerene molecule, respectively; 
$\epsilon_{ij}$ and $\sigma _{ij}$ are the mixed LJ potential parameters, obtained from the 
usual Lorentz-Berthelot combination rules $\epsilon _{ij}=\sqrt{\epsilon _{ii}\epsilon _{jj}}$ 
and $2\sigma_{ij}=\sigma _{ii}+\sigma _{jj}$ and $\textbf{r}_{ij}$ is the distance vector 
between the \textit{i}th atom of the molecule and the \textit{j}th carbon atom.

In the second term of Eq. (\ref{eq1}) $\textbf{E}_{\text{M}}^j$ is the field generated by 
the molecular permanent electric multipole moments ($\mathbf{\mu }$, $\mathbf{\Theta }$,...) 
of the NH$_3$ molecule at position \textbf{r}$_0$ on the \textit{j}th carbon atom of the 
fullerene molecule at position \textbf{r}$_j$ with polarizability 
tensor $\mathbf{\alpha }^j$. It is expressed as

\begin{eqnarray}
\textbf{E}_{\text{M}}^j = \mathbf{\nabla\nabla }\left( \frac 1{\left| \mathbf{r}_j-\mathbf{r}_0\right|
}\right) \cdotp\mathbf{\mu} + \frac 13 \mathbf{\nabla\nabla
\nabla }\left( \frac 1{\left| \mathbf{r}_j-\mathbf{r}_0\right| }\right)
\mathbf{:\Theta }+... \label{eq2}
\end{eqnarray}

Note that the multipole moments of NH$_3$ and the polarizability tensor of each carbon atom 
of C$_{60}$ are usually defined and given with respect to local frames with 
their origins at the centre of mass of NH$_3$ (G,\textbf{x},\textbf{y},\textbf{z}) or 
each carbon atom (C$_j$,\textbf{x}$_j$,\textbf{y}$_j$,\textbf{z}$_j$) respectively. 
In the calculation, it is then necessary to express these quantities in an absolute 
frame (O,\textbf{X},\textbf{Y},\textbf{Z}) defined with respect to the fixed C$_{60}$ 
molecule. We use a transformation rotation matrix following Rose's 
convention \cite{mer67} and given in Appendix B.

Figure 1 indicates how to describe the internal degrees of freedom of the
molecule with respect to its frame (G,\textbf{x},\textbf{y},\textbf{z}) 
and its external orientational and translational degrees of freedom with respect 
to the absolute frame, and Table 1 gives the various fullerene 
and ammonia characteristic values used in our calculations.

\begin{figure}[h!]
\begin{center}
\includegraphics[width=16cm]{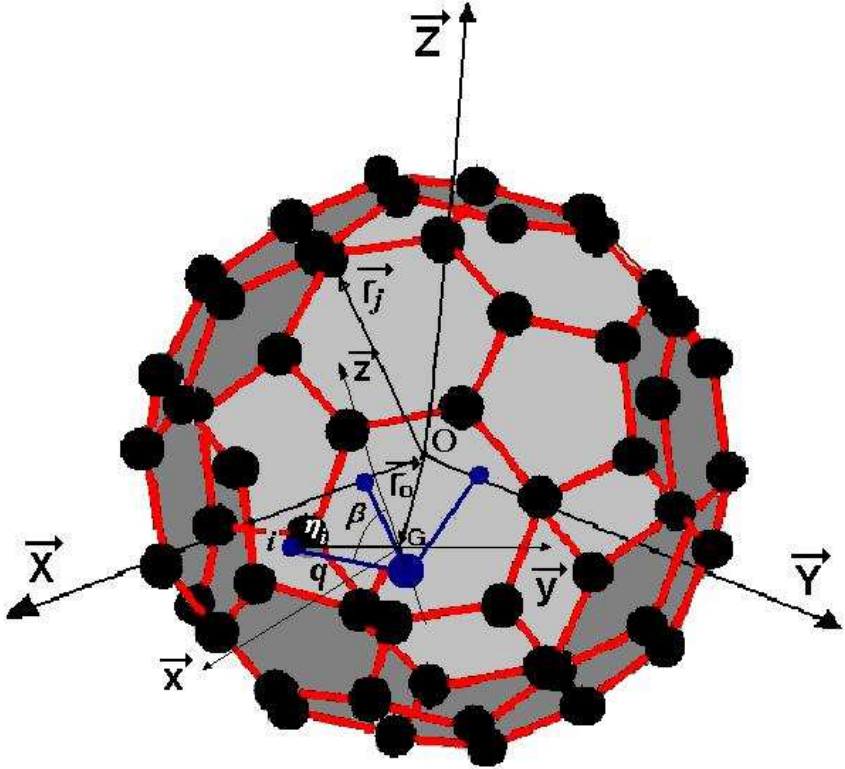}
\end{center}
\caption{Geometrical characteristics of a molecule trapped in the
fullerene molecule. (O,\textbf{X},\textbf{Y},\textbf{Z}) 
and (G,\textbf{x},\textbf{y},\textbf{z}) represent the absolute 
frame and the molecular frame, respectively. \textit{q} 
and $\beta$ define the equilibrium internal coordinates of the trapped molecule.} 
\label{FIG. 1}
\end{figure}

\begin{table}[h!]
\begin{center}
\caption{Numerical parameters for NH$_3$ molecule and 
for C$_{60}$ molecule used in our calculations.}
\begin{tabular}{llll}
\hline\hline
NH$_3$ & &  C$_{60}$  & \\
\hline
$q$ (\AA ) & 1.0156 &  $R^{(a)}$ (\AA ) & 3.556 \\
$\beta$ (deg.) & 68. &  d(C$-$C)$^{(a)}$ (\AA ) & 1.455 \\
 &  & d(C$=$C)$^{(a)}$ (\AA ) & 1.398 \\
$\mu ^{\mathrm{e}}$ (D) & 1.476 &  $\alpha _{\bot }^{\mathrm{c}}$
(\AA $^3$) & 1.44 \\
$\Theta ^{\mathrm{e}}$ (D\AA ) & -2.930 &  $\alpha _{\Vert }^{\mathrm{c}%
}$ (\AA $^3$) & 0.41 \\
$B$ (cm$^{-1}$) & 9.941 & &  \\
$C$ (cm$^{-1}$) & 6.309 & &  \\
\hline
\hline

 & C - C$^{(b)}$ &  N - N$^{(c)}$ & H - H$^{(c)}$   \\
\hline
$\epsilon$ (cm$^{-1}$) & 19.4 & 26.5 & 17.2 \\
$\sigma$ (\AA ) & 3.40 & 3.38 & 2.53 \\

\hline

\raggedright $(a)$ From Ref. \cite{khlhdbcabhcdrdjmdv91}.
\raggedright $(b)$ From Ref. \cite{hysxwz2009}.
\raggedright $(c)$ From Ref. \cite{al2000}.

\end{tabular}
\end{center}
\label{Table 1}
\end{table}

The distance vector $\mathbf{r}_{ij}$ in Eq.(\ref{eq1}) can be expressed in 
terms of the position vectors \textbf{r}$_0$ of the molecular centre of mass (c.m.) 
and $\mathbf{r}_j$ of the $j$th carbon atom of the fullerene molecule with respect 
to the absolute frame (O,\textbf{X},\textbf{Y},\textbf{Z}) and of the position 
vector $\mathbf{\eta }_i$ of the $i$th atom of the molecule with respect to its 
associated frame (G,\textbf{x},\textbf{y},\textbf{z}):

\begin{equation}
\mathbf{r}_{ij} = \mathbf{r}_j-\mathbf{r}_0-\mathbf{\eta }_i, \label{eq3}
\end{equation}

The IR spectra of the trapped NH$_3$ molecule are usually given in terms of 
molecular vibrational \{$Q$\} degrees of freedom.

Thus, to determine the influence of the surroundings on the molecular
internal motions, the position vectors $\mathbf{\eta }_i$ of the atoms in the
molecule, the molecular dipole moment vector $\mathbf{\mu }$, and the
quadrupole moment tensor $\mathbf{\Theta }$ can all be expressed in a series
expansion with respect to the molecular frame in terms of the vibrational
normal coordinates \{$Q$\}.

Taking only the first order terms gives

\begin{eqnarray}
\mathbf{\eta }_i &=&\mathbf{\eta }_i^{\text{e}}+\sum\limits_\nu \mathbf{a}%
_i^\nu Q_\nu , \nonumber \\
\mathbf{\mu } &=&\mathbf{\mu }^{\text{e}}+\sum\limits_\nu \mathbf{b}^\nu
Q_\nu , \nonumber \\
\mathbf{\Theta } &=&\mathbf{\Theta }^{\text{e}}+\sum\limits_\nu \mathbf{c}%
^\nu Q_\nu . \label{eq4}
\end{eqnarray}

In these expressions $\mathbf{\eta }_i^{\text{e}}$ is the vector position of 
the \textit{i}th atom in the rigid molecule, and $\mathbf{\mu }^{\text{e}}$ 
and $\mathbf{\Theta }^{\text{e}}$ are its permanent multipole moments, 
while $\mathbf{a}_i^\nu $, $\mathbf{b}^\nu $, and $\mathbf{c}^\nu $ are the 
first derivatives of $\mathbf{\eta }_i$, $\mathbf{\mu }$, and $\mathbf{\Theta }$ 
with respect to the normal coordinate $Q_\nu $ which describes the $\nu $th 
molecular vibrational mode with frequency $\omega _\nu $. 
Note that the expression of $\mathbf{\mu }$ will also be used in the near and 
far infrared spectra calculations.

However, it must be noticed that because of the \textit{dual nature} of the 
$\nu_2$ mode of the ammonia molecule, that is, the \textit{high frequency vibration-low 
frequency inversion}, this mode will be treated in a particular way. 

Using the Born-Oppenheimer approximation (adiabatic approximation) to separate the 
high frequency vibrational modes \{$Q$\} of the molecule from its low frequency 
external modes, that is, the orientation $\mathbf{\Omega}=(\varphi,\theta,\chi)$ and 
c.m. translation \textbf{r}$_0$ motions, the interaction 
potential energy $V_{\text{MC}}$ can be written as 

\begin{equation}
V_{\text{MC}} = V_{\text{M}}(\mathbf{r}_0,\mathbf{\Omega}) + 
W_{\text{M}}(\{Q\}) + \bigtriangleup V_{\text{M}}(\mathbf{r}_0,\mathbf{\Omega},
\{Q\}), \label{eq5}
\end{equation}

\noindent where $V_{\text{M}}(\mathbf{r}_0,\mathbf{\Omega})$ represents the low frequency 
motions dependent part of the potential energy for the non vibrating molecule, 
and $W_{\text{M}}(\{Q\})$ characterizes the vibrational dependent part for 
the molecule at its equilibrium position and orientation, and is generally a small 
perturbation which involves only vibrational level shifts and/or splittings. 
It can be developed in a Taylor series expansion up to second order with respect to 
the vibrational normal coordinates \{$Q$\} as

\begin{equation}
W_{\text{M}}(\{Q\}) = \sum\limits_\nu \frac{\partial
V_{\text{MC}}}{\partial Q_\nu }Q_\nu +\frac 12\sum\limits_{\nu \nu ^{\prime }}\frac{
\partial ^2V_{\text{MC}}}{\partial Q_\nu \partial Q_{\nu ^{\prime }}}Q_\nu Q_{\nu
^{\prime }}+... \label{eq6}
\end{equation}

Finally the third term in Eq.(\ref{eq5}) characterizes the coupling between the 
external and the internal modes which can lead to the vibrational energy relaxation.

\subsection{Potential energy surfaces}

\subsubsection{Orientation-translation motions}

The first step of the numerical calculations consists in determining the equilibrium 
configuration of the rigid ammonia molecule into the C$_{60}$ cage. The potential 
energy $V_{\text{M}}$ (see Eq.(\ref{eq5}) is minimized with 
respect to both the c.m. displacement vector \textbf{r}$_0$ and 
the molecular orientional coordinates $\varphi$, $\theta$, and $\chi$. 

We find that the equilibrium configuration has an energy minimum value of 
- 1460 cm$^{-1}$ for the \textbf{z} molecular axis (threefold symetry axis C$_3$) 
oriented along the direction connected to each two 
symmetrical carbon atoms. The c.m. is then displaced from the site centre by 
about 0.184 $\text{\AA}$ in the same direction. 
An energy maximum of - 1445 cm$^{-1}$ is obtained when the molecule is oriented and 
its c.m. displaced by 0.188 $\text{\AA}$ along the pentagon centre directions. 
There is also an intermediate energy value of - 1455 cm$^{-1}$ that is calculated for 
the molecule oriented and its c.m. displaced by 0.180 $\text{\AA}$ along the hexagon 
centre directions. 

Thus, the calculations show that \textit{i}) in all the molecular orientational 
configurations, the c.m. displacements \textit{r}$_0$ have always opposite directions 
to the molecular \textbf{z} axis, and \textit{ii}) the energy changes are negligibly 
small $\lesssim$ 3 cm$^{-1}$ when \textit{r}$_0$ changes by about 0.01 $\text{\AA}$ 
along these directions. So, in the following we will take the value 
\textit{r}$_{0}^{\text{e}}$ = 0.184 $\text{\AA}$ as a mean value for every orientation. 
This means that the trajectory of the molecular c.m. can be approximated by the 
surface of a sphere of radius \textit{r}$_{0}^{\text{e}}$. 

Note, moreover, that the induction and repulsion-dispersion contributions 
represent, respectively, 35$\%$ (- 498 cm$^{-1}$) and 65$\%$ (- 963 cm$^{-1}$) of 
the total energy.

As a result, the potential energy surface 
$V_{\text{M}}(\mathit{r}_0(\mathbf{\Omega}))$ is nearly flat since the barrier 
height is always $\lesssim$ 16 cm$^{-1}$ and the c.m. displacement vectors \textbf{r}$_0$ 
from the cage centre remain always nearly anticollinear to the molecular \textbf{z} 
axis \textit{i.e.} \textbf{z}$\cdotp$ \textbf{r}$_0$ $\simeq$ - \textit{r}$_0$. 
This is not surprising because of the repulsive character of the hydrogen atoms 
of the molecule and reflects a very strong coupling between its orientation 
and the location of its centre of mass. 

Thus, on one hand, the axis of the molecule undergoes only oscillatory motions 
($\varphi,\theta$) of small amplitude around the direction of each displacement vector 
\textbf{r}$_{0}^{\text{e}}$. The corresponding potential energy surface which is given in 
Figure 2 is a nearly harmonic two dimension librational oscillator. 
The associated frequencies are $\omega_{\varphi}$ $\simeq$ 155 cm$^{-1}$ and 
$\omega_{\theta}$ $\simeq$ 151 cm$^{-1}$.

\begin{figure}[h!]
\begin{center}
\includegraphics[width=10cm]{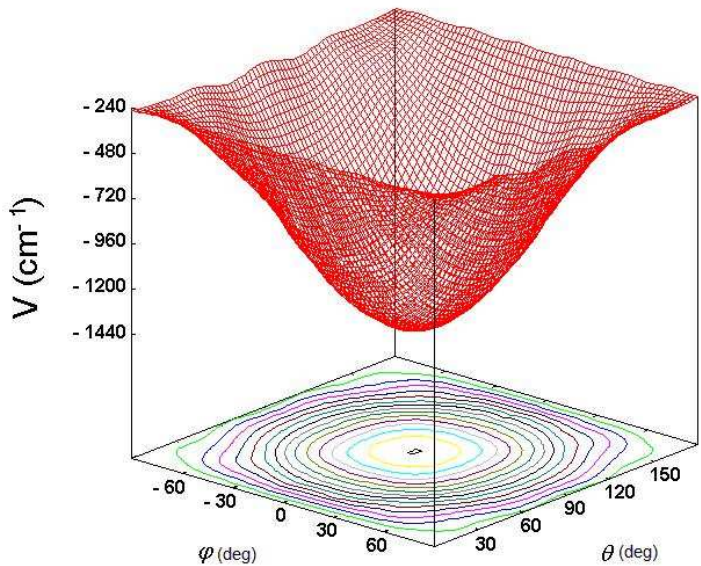}
\end{center}
\caption{Potential energy surface \textit{versus} ($\varphi,\theta$) angular motions 
around the molecular equilibrium configuration 
($\mathbf{r}_{0}^{\text{e}}$,$\mathbf{\Omega}$).}
\label{FIG. 2}
\end{figure}

On the other hand, for a given molecular orientation, the radial potential 
function $V_{\text{M}}(\textit{r}_0)$ of the translation motion of the molecular c.m. 
around the equilibrium position $r_{0}^{\text{e}}$ = 0.184 $\text{\AA}$ is 
plotted in Figure 3. The curve appears as that of a nearly harmonic one dimensional 
oscillator with a force 
constant value \textit{k} = 2.4$\times$10$^4$ cm$^{-1} \ldotp \text{\AA}^{-2}$ 
corresponding to a frequency $\omega_{\text{rad}}$ $\simeq$ 218 cm$^{-1}$. 

\begin{figure}[h!]
\begin{center}
\includegraphics[width=10cm]{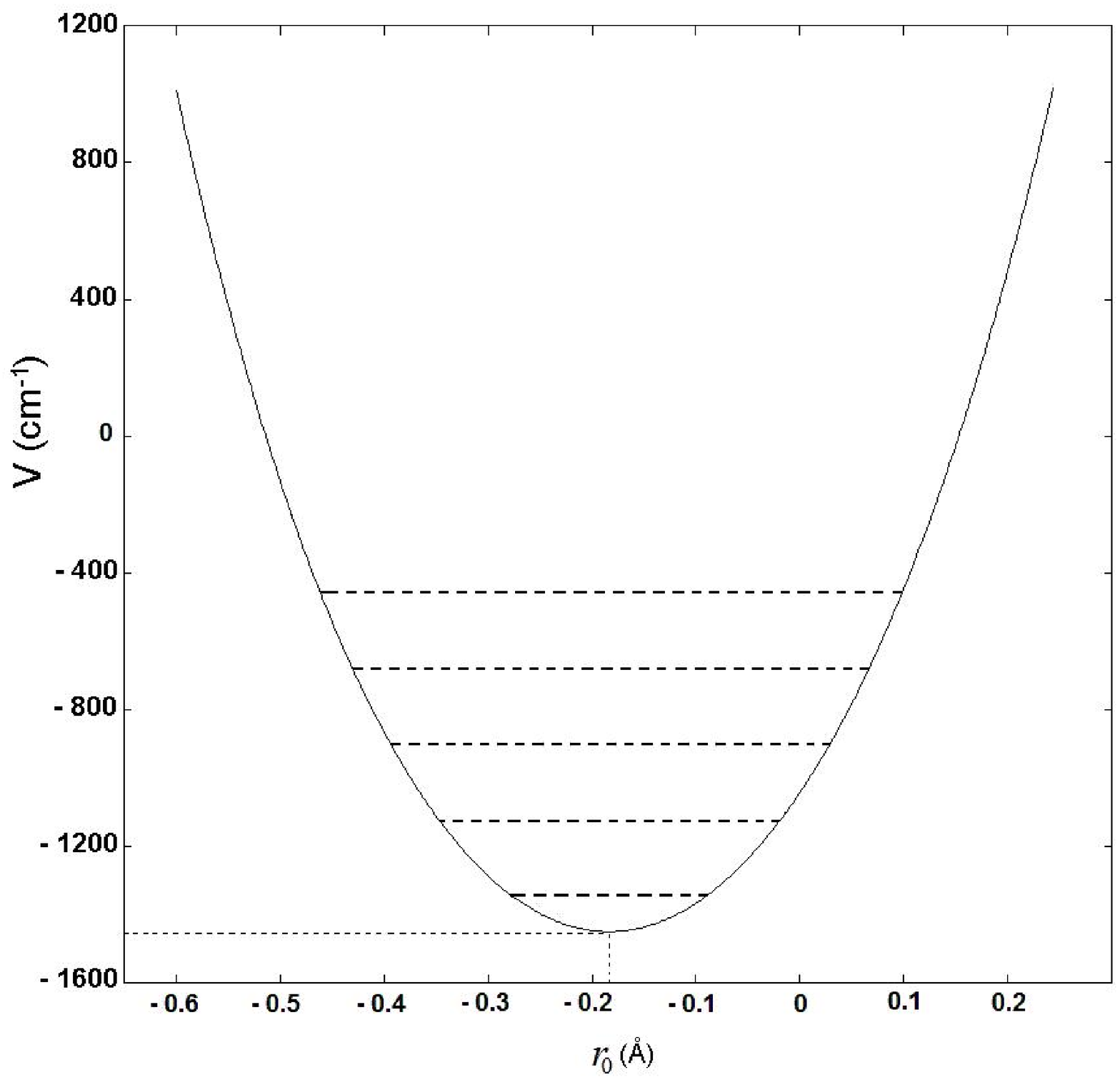}
\end{center}
\caption{Potential energy function \textit{versus} 
molecular c.m. displacement $r_0$ around its equilibrium position 
vector \textbf{r}$_{0}^{\text{e}}$.}
\label{FIG. 3}
\end{figure}

On the basis of these features, we can conclude that ammonia should exhibit a 
nearly free rotation motion simultaneously around the centre of the cage by describing 
a spherical curve with radius \textit{r}$_{0}^{\text{e}}$ = 0.184 $\text{\AA}$, and 
around its centre of mass. Note however that the spinning motion $\chi$ remains around 
the threefold symmetry axis of the molecule, only.

Finally, it is clear that this rotation-c.m. translation motion is slow with respect to the 
librational and the radial motions and can be separated from the latter using the 
adiabatic approximation.

\subsubsection{Inversion-translation motions}

It is also clear, from the results above, that when the inversion motion of the molecule 
proceeds, its centre of mass moves simultaneously from one position 
(\textbf{r}$_{0}^{\text{e}},\mathbf{\Omega}$) to the opposite one 
(-\textbf{r}$_{0}^{\text{e}},\mathbf{\Omega}$) with respect to the cage 
centre where the molecule lies in its planar configuration. 

\subsubsection{Few remarks}

It must be noticed that the interaction potential energy experienced by the ammonia 
molecule trapped in the fullerene molecule strongly depends on the geometrical 
characteristics such as the radius of the C$_{60}$ cage and the equilibrium 
internal coordinates (\textit{q},$\beta$ see Fig. 1) of the ammonia molecule, and 
also on the dispersion-repulsion potential parameters especially the 
Lennard-Jones $\sigma$ parameters. 

For instance, in one hand, the minimum depth is enhanced from - 1460 
to - 750 cm$^{-1}$ when the fullerene radius is decreased 
by 1$\%$, from 3.556 to 3.519 $\text{\AA}$, and it diminishes 
to - 2000 and - 2411 cm$^{-1}$ when the 
radius value is increased by 1$\%$ and 2$\%$, to 3.592 and 3.627 $\text{\AA}$, 
respectively. The induction energy contribution undergoing only weakly variations. 

In the other hand, changes of the $\sigma$ parameters of + 1$\%$ and - 1$\%$ 
involve, respectively, minimum depths of - 952 and - 1976 cm$^{-1}$.

However, in spite of these changes in the well depth values, the potential energy surface 
$V_{\text{M}}(\mathbf{r}_{0}^{\text{e}},\mathbf{\Omega})$ remains nearly flat since 
its barrier height does not exceed 24 cm$^{-1}$. Furthermore, the potential energy surfaces 
associated with the various motions remain nearly unchanged. 

Thus, from the previous analysis of the potential energy experienced by the 
ammonia molecule into the fullerene molecule, we will assume that the c.m. position 
vector \textbf{r}$_{0}$ remains permanently anticollinear to the \textbf{z} molecular 
axis and separate it into a static position vector 
\textbf{r}$_{0}^{\text{e}}(\varphi,\theta,Q_2)$ which parametrically depends on 
the orientation and inversion degrees of freedom, and a dynamical 
coordinate \textbf{u}$_{0}$ characterizing the c.m. vibration 
around \textbf{r}$_{0}^{\text{e}}(\varphi,\theta,Q_2)$. 

Finally, it is clear that inversion-c.m. translation motions 
(\textit{r}$_{0}^{\text{e}}$,$Q_2$), on one hand, and 
orientation-c.m. translation motions (\textbf{r}$_{0}^{\text{e}}$,$\mathbf{\Omega}$), 
on the other hand, are strongly coupled; their kinetic operator must be classically 
treated, as explained in appendix A, before their quantum mechanical treatment as 
described below. The interaction potential energy $V_{\text{MC}}$ can be rewritten 
as 

\begin{equation}
V_{\text{MC}} = V_{\text{M}}(\mathbf{r}_{0}^{\text{e}},\mathbf{\Omega}) + 
V_{\text{M}}(\textit{r}_{0}) + V_{\text{M}}(\textit{r}_{0},Q_2) + 
W_{\text{M}}(\{Q\}) + \bigtriangleup V_{\text{M}}(\mathbf{r}_0,\mathbf{\Omega},
\{Q\}), \label{eq7}
\end{equation}

\section{Quantum-classical model}

As we have mentioned above, the purpose of this work is to determine the infrared 
spectra of the NH$_3$ molecule trapped in a rigid C$_{60}$ molecule. The Hamiltonian 
of the optically active molecule considered as quantum mechanical system can 
be expressed from the potential energy surfaces calculated above and the quantum 
mechanical model for the orientational and vibrational kinetic energy operators. 
The latter are obtained from the correspondence principle applied to the classical 
formulation described in Appendix A. Disregarding the dynamical couplings between 
the orientational and vibrational degrees of freedom the Hamiltonian can thus be 
written as 

\begin{equation}
H_{\text{a}}^{\text{eff}} = H_{\text{orient}}^{\text{eff}} + H_{\text{vib}}^{\text{eff}};
\label{eq8}
\end{equation}

\noindent where $H_{\text{orient}}^{\text{eff}}$ and $H_{\text{vib}}^{\text{eff}}$ 
are, respectively, the renormalized Hamiltonians associated with the 
orientation and the vibration modes which account for the couplings 
(\textbf{r}$_{0}^{\text{e}}$,$\mathbf{\Omega}$) 
and (\textit{r}$_{0}^{\text{e}}$,$Q_2$).

\subsection{The orientation modes}

Following expressions derived in Appendix A, the 
quantum mechanical renormalized Hamiltonian $H_{\text{orient}}^{\text{eff}}$ for 
the non-vibrating molecule can be written as 

\begin{equation}
H_{\text{orient}}^{\text{eff}} = T_{\text{rot}}^{\text{eff}} + 
V_{\text{M}}(\mathbf{r}_{0}^{\text{e}},\mathbf{\Omega}); \label{eq9}
\end{equation}

\noindent where $T_{\text{rot}}^{\text{eff}}$ is 
the effective rotational kinectic operator 

\begin{eqnarray}
T_{\text{rot}}^{\text{eff}} = -B^{\text{eff}}\left\{ \frac{\partial ^2}{%
\partial \theta ^2} + \cot \theta \frac
\partial {\partial \theta } + \frac 1{\sin ^2\theta }\frac{\partial ^2}{%
\partial \varphi ^2} + \left( \cot ^2\theta +
\frac {C}{B^{\text{eff}}}\right) \frac{\partial ^2%
}{\partial \chi ^2} - 2\frac{\cot \theta }{\sin \theta }\frac{\partial ^2}{%
\partial \varphi \partial \chi }\right\}. \label{eq10}
\end{eqnarray}

In this equation $B^{\text{eff}}$ = $\hbar^2/2I_{\text{B}}^{\text{eff}}$ is 
the effective rotational constant connected to the rotation-c.m. translation motion of the 
C$_3$ symmetry axis for the trapped molecule and $C$ its rotational constant of 
the spinning motion around this axis for the gas-phase molecule 
$C$ = 6.309 cm$^{-1}$. The value of $B^{\text{eff}}$ is equal 
to 7.419 cm$^{-1}$ which is less than the gas-phase value of 9.941 cm$^{-1}$.

Furthermore, since the barrier height associated with the orientation-c.m. translation 
motion is always $\lesssim$ 16 cm$^{-1}$, the potential 
$V_{\text{M}}(\mathbf{r}_{0}^{\text{e}},\mathbf{\Omega})$ can be neglected. 
The renormalized Hamiltonian becomes 
$H_{\text{orient}}^{\text{eff}} \simeq T_{\text{rot}}^{\text{eff}}$.

The eigensolutions of this Hamiltonian are then 
$E_{JMK}=B^{\text{eff}}J(J+1)+(C-B^{\text{eff}})K^2$ for the eigenvalues 
(energy levels) and $|JMK\rangle$ for the eigenvectors where $J$, $M$, and $K$ 
are the usual quantum numbers connected to the free rotation motion of the 
ammonia molecule with the conditions $-J\leq M \leq +J$ and $-J\leq K \leq +J$. 
Note that these levels have (2$J$+1)-fold degeneracy on the quantum 
number $M$ and twofold degeneracy ($\pm$) on the $K$ one, except when $K=0$ \cite{chtals55}.
The corresponding line structure of the far infrared (FIR) bar-spectrum is then 
identical to that of NH$_3$ gas-phase one.

\subsection{The $\nu_2$ vibration-inversion mode}

\subsubsection{Gas-phase NH$_3$}

In the vibration-inversion mode of the NH$_3$ molecule, the nitrogen atom can pass 
from one side of the plane defined by the three hydrogen atoms to the opposite one. 
This mode is generally described, to a good approximation, as one-dimension 
tunneling motion of one particle of mass equal 
to the reduced mass $\mu_2$ of NH$_3$ (2.563 g.mol$^{-1}$) moving in a symmetrical 
double-well potential function $V_{\nu_2}(Q_2)$ (left (L) and right (R) wells) as 
shown in Figure 4a, with a high but finite hindering barrier 
of 2047 cm$^{-1}$ \cite{jdsjai62}, such as

\begin{equation}
V_{\nu_2}(Q_2) = \frac 12 k_2 Q_{2}^2 + A_2 (\text{e}^{-a_{2}Q_{2}^2} - 1); \label{eq11} 
\end{equation}

\noindent where $k_2$ = 58306 cm$^{-1} \ldotp \text{\AA}^{-2}$, $A_2$ = 12469 cm$^{-1}$ and 
$a_{2}$ = 4.8224 $\text{\AA}^{-2}$.

The solution of such a problem leads to vibrational energy levels split into 
doublets, ($+$) and ($-$), due to inversion. The vibration-inversion wavefunctions 
are then described by $|v_{2}^{(\alpha)}\rangle$ ($\alpha = +,-$), instead of 
$|v_2\rangle$, and are expressed as linear combinations (symmetric and 
antisymmetric) of the vibrational wavefunctions $|v_{2}^{(\text{L})}\rangle$ and 
$|v_{2}^{(\text{R})}\rangle$ associated with the left and right wells, respectively. 
Moreover, the only permitted transitions are $(+) \longleftrightarrow (-)$ as shown 
in Fig. 4a.

Note that the potential parameters given above were adjusted to give the fundamental 
frequency and the level splittings corresponding to the experimental 
values, that is, 950 cm$^{-1}$ for the frequency 
and 0.8 cm$^{-1}$ for $v_2$ = 0 and 35.8 cm$^{-1}$ for $v_2$ = 1 level splittings. 

\subsubsection{Trapped NH$_3$}

When the NH$_3$ molecule is trapped in C$_{60}$ cage, the gas-phase double-well 
potential function is modified. In the calculations, the inversion-c.m. translation 
potential energy $V_{\text{M}}(r_0,Q_2)$ (see Eq. (\ref{eq7})) is numerically 
added to the potential function $V_{\nu_2}(Q_2)$ of Eq. (\ref{eq11}). 
The new double-well potential function is shown in Fig. 4b as a function of 
the inversion angular coordinate. Its symmetry is preserved but the hindering 
barrier is increased by 404 cm$^{-1}$. 
Moreover, from expressions given in Appendix A, the effective reduced 
mass is equal to $\mu_{2}^{\text{eff}}$ = 6.569 g.mol$^{-1}$ leading to 
an increase of 156$\%$ with respect to the gas-phase value of 2.563 g.mol$^{-1}$.

The corresponding Schr\"{o}dinger equation was solved numerically using 
a discrete variable representation method \cite{jcliphjvl85}. The computed level 
scheme is given in Fig. 4b. 

In gas-phase, the $\nu_2$ mode is measured at 931.7 cm$^{-1}$ and 968.3 cm$^{-1}$. 
In C$_{60}$, it is shifted by 47.0 cm$^{-1}$ (as given below) and because the 
double-well inversion potential is modified with an increase in the hindering 
barrier height and the effective reduced mass as discussed above, this 
mode is quasi-degenerate and is calculated at 751.6 cm$^{-1}$, that is, 
(704.6+47.0) cm$^{-1}$ (see Fig. 4b). This result is to be related to an 
increase in the tunneling 
time for both the fundamental and the first excited levels in C$_{60}$, since 
the values are equal to 20.85 ps and 0.47 ps in gas-phase compared 
to 55594 ps and 333.56 ps in C$_{60}$, respectively. We can infer from these 
results that in C$_{60}$, localization of the N atom occurs in one of the 
wells of the potential function and that the $\nu_2$ mode is quasi-degenerate.

\begin{figure}[h!]
\begin{center}
\includegraphics[width=7cm]{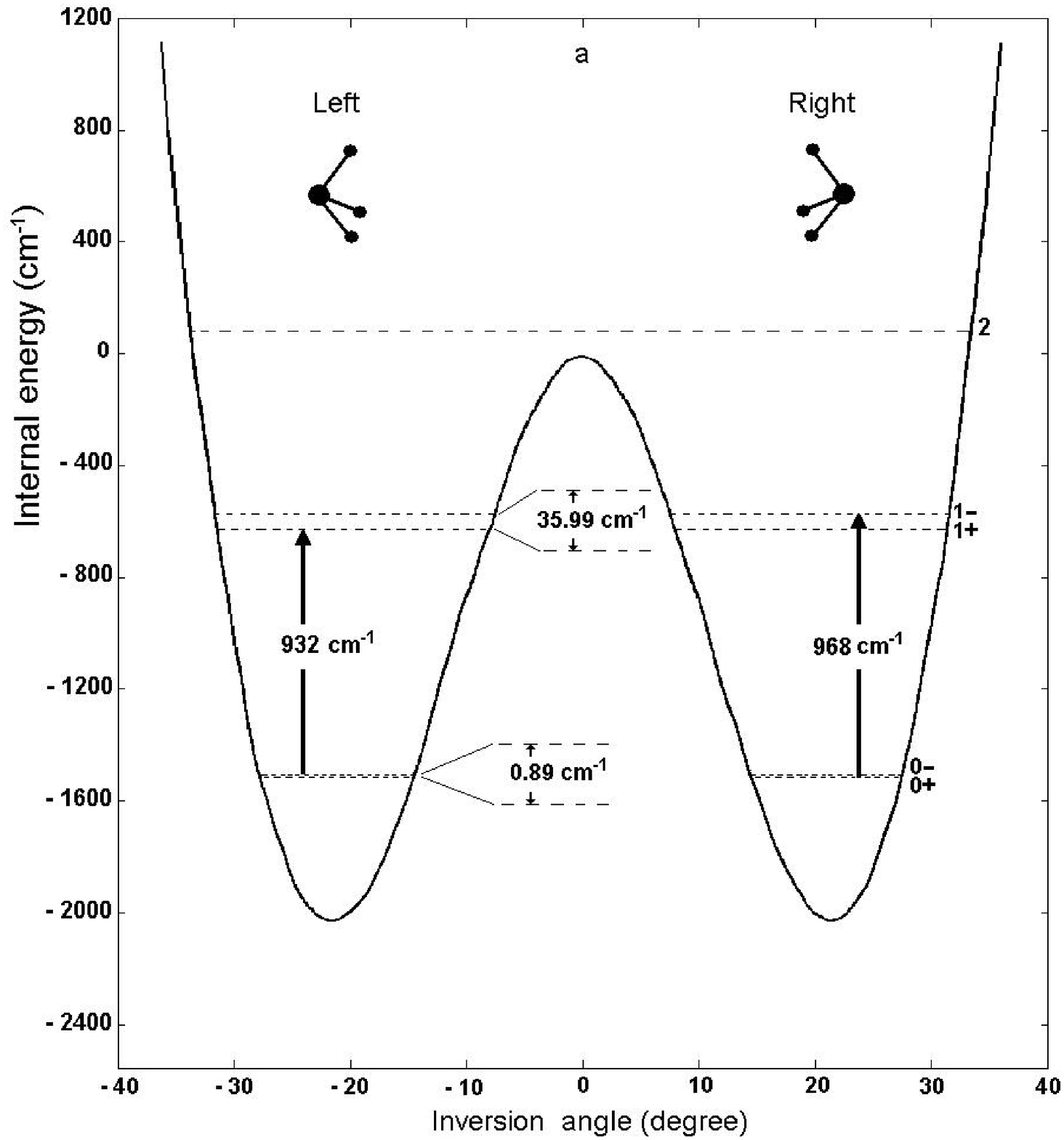}
\includegraphics[width=7cm]{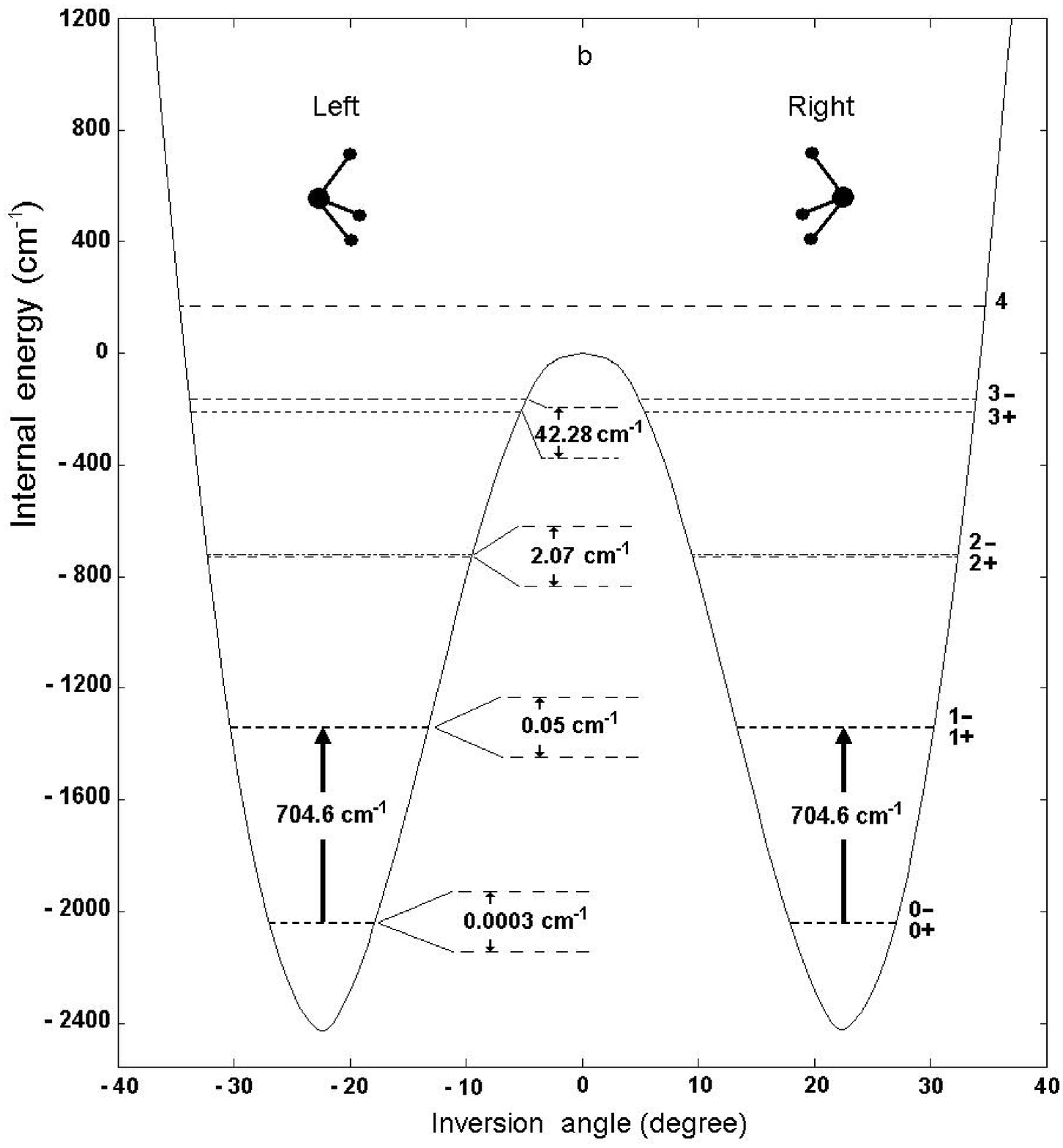}
\end{center}
\caption{Vibration-inversion double-well potential energy of NH$_3$ as a function of 
the angle between the N$-$H bond and the plane of the H atoms. The obtained energy 
levels and the allowed fundamental transitions are also shown (a) for the 
gas-phase and (b) for the trapped one in the fullerene nano-cage.} 
\label{FIG. 4}
\end{figure}

\subsection{The vibrational frequency shifts}

The second term of Eq. (\ref{eq8}) is the renormalized vibrational Hamiltonian 
which accounts for the vibrational dependent part 
$W_{\text{M}}(\{Q\})$ and inversion-c.m. translation part $V_{\text{M}}(r_0,Q_2)$ of the 
interaction potential energy 

\begin{equation}
H_{\text{vib}}^{\text{eff}} = H_{\text{vib}} + W_{\text{M}}(\{Q\}) + V_{\text{M}}(r_0,Q_2); 
\label{eq12}
\end{equation}

\noindent where $H_\text{vib}$ characterizes the gas-phase molecular vibrational 
Hamiltonian written as 

\begin{equation}
H_{\text{vib}} = \sum\limits_\nu \frac{P_\nu ^2}{2\mu _\nu }+\frac 12 \sum\limits_\nu k_\nu
Q_\nu ^2+ \bigtriangleup V_{\text{vib}}^{\text{anh}}(\{Q\}). \label{eq13}
\end{equation}

In this expression $\mu _\nu$ and $k _\nu$ are, respectively, the reduced mass and 
the harmonic force constant connected to the $\nu$th vibrational mode, with associated 
normal coordinate $Q_\nu$ and conjugate momentum $P_\nu$, and 
$\bigtriangleup V_{\text{vib}}^{\text{anh}}(\{Q\})$ represents the anharmonic part of 
the internal potential function. The eigenelements of this Hamiltonian $H_{\text{vib}}$ 
have previously been obtained from the \textit{ab initio} calculations of 
Martin \textit{et al.} \cite{jmlmtjlprt92}. 

Let $E_{v_\nu}$ and $|...v_\nu...\rangle$ be the eigenvalues and eigenvectors of 
the \textit{v}th level associated with the $\nu$th vibrational mode. 

Thus, a first order perturbation treatment allows us to determine the frequency 
shift $\Delta \omega _\nu$ associated with the vibrational fundamental transition 
$\mid...0_\nu ...\rangle \rightarrow \mid ...1_\nu...\rangle $ for 
the trapped molecule from the equation 

\begin{equation}
\Delta \omega _\nu = \hslash^{-1}\left[ \langle ...1_\nu...\mid W_{\text{M}}(\{Q\})
\mid ...1_\nu...\rangle - 
\langle ...0_\nu ...\mid W_{\text{M}}(\{Q\})\mid ...0_\nu...\rangle \right]; \label{eq14}
\end{equation}

\noindent in which all other modes remain in their fundamental states 
and $\hslash = h/2\pi$. The computed values are given in Table 3. We note that each 
of the vibrational modes are blue shifted with splitting of the degenerate 
levels of $\nu_3$ mode (strong) and $\nu_4$ mode (weak) modelled from static effect. 
Whereas for degenerate $\nu_2$ mode the blue shift of 47.0 cm$^{-1}$ from static 
effect, is modified to red shift when dynamic effect on 1D tunneling motion 
of the N atom is considered as discussed above.

\begin{table}[h!]
\caption{Vibrational frequency shifts (cm$^{-1}$) of NH$_3$ molecule 
trapped in the C$_{60}$ cage, together with the gas-phase frequencies. }
\begin{tabular}{ccc}
\hline\hline
\begin{tabular}{c} Vibrational modes \end{tabular} & \begin{tabular}{c} Frequency 
shifts (cm$^{-1}$) \end{tabular} & \begin{tabular}{c} Gas-phase  frequencies (cm$^{-1}$)
\end{tabular} \\
\hline
$\nu_1$ & \multicolumn{1}{c}{133.0} & \multicolumn{1}{c}{3337} \\
$\nu_2$ & \multicolumn{1}{c}{47.0} & \multicolumn{1}{c}{950} \\
$\nu_{3a}$ & \multicolumn{1}{c}{171.0} & \multicolumn{1}{c}{3448} \\
$\nu_{3b}$ & \multicolumn{1}{c}{69.6} & \multicolumn{1}{c}{} \\
$\nu_{4a}$ & \multicolumn{1}{c}{47.1} & \multicolumn{1}{c}{1627} \\
$\nu_{4b}$ & \multicolumn{1}{c}{34.0} & \multicolumn{1}{c}{} \\ 
\hline
\end{tabular}
\label{Table 2}
\end{table}

\section{Infrared absorption spectra}

\subsection{General}

The infrared absorption coefficient for the optically active ammonia molecule 
trapped in a C$_{60}$ fullerene molecule at temperature \textit{T}, is defined 
as the real part of the spectral density, \textit{i.e.}, the Fourier transform of the
time-dependent autocorrelation function $\phi (t)$

\begin{eqnarray}
I\left( \omega \right) &=&\frac{4\pi \mathcal{N}\omega }{3hc}\mathrm{Re}%
\int\limits_0^\infty dte^{i\omega t}\phi (t), \nonumber \\
\phi (t) &=&\text{Tr}\left[ \rho (0)\mathbf{\mu }_A(0)\mathbf{\mu }%
_A(t)\right], \label{eq15}
\end{eqnarray}

\noindent where $\omega $ is a running frequency variable
and $c$ the speed of light, $\rho (0)$ 
characterizes the initial canonical density matrix operator of the system, and 
$\mu _A$ is the molecular dipole moment operator defined in the absolute frame 
(see Appendix B). The trace (Tr) operation means an average over the initial 
conditions at time $t=0$ and over all the possible evolutions of the system
between times $0$ and $t$.

However, since this work is devoted to constructing the infrared bar-spectra 
of the trapped ammonia molecule, we assume the \textit{initial chaos hypothesis} for the 
density matrix to be valid. This allows us to write the optical wave functions 
$\mid v_\nu JMK\rangle$ as products of the renormalized vibrational and 
orientational ones $\mid v_\nu \rangle \otimes |JMK\rangle_{v_{\nu}}$. 
It must be noticed that the orientational states parametrically depend on the 
vibrational states through the moments of inertia. This dependence will be 
ignored below.

\subsection{Near infrared bar-spectrum}

Within this framework, the infrared bar-spectrum connected to the $\nu$th 
fundamental vibrational mode transition 
$\mid 0_\nu\rangle \longrightarrow \mid 1_\nu\rangle $ 
(all other modes being in their fundamental states) is written in the Lorentzian form as

\begin{eqnarray}
I_\nu (\omega ) = \frac{8\pi ^2\mathcal{N}}{3hc}\omega
\left| \left\langle 0_\nu \left| Q_\nu \right|
1_\nu\right\rangle \right| ^2 
\sum\limits_{ {JMK}{J^{\prime }M^{\prime }K^{\prime }}}\frac{%
e^{-\beta E_{0JMK}}-e^{-\beta E_{1J^{\prime }M^{\prime }K^{\prime}}}}Z 
\left| \left\langle JMK\left| \frac{\partial \mathbf{\mu }_A%
}{\partial Q_\nu }\right| J^{\prime }M^{\prime }K^{\prime }%
\right\rangle \right| ^2 \nonumber\\
\times \delta \left(\omega -\omega _{0JMK\rightarrow
1J^{\prime }M^{\prime }K^{\prime }}-\Delta \omega _{\nu}\right). 
\label{eq16}
\end{eqnarray}

\noindent In this equation the $\left\langle ...\right\rangle $ brackets
refer to vibrational transition elements of the normal coordinate $Q_\nu $
and to the orientational transition elements of the first derivative of the
molecular dipole moment with respect to this coordinate. The numerical values 
connected to all the NH$_3$ modes are given in Appendix B. Moreover, $Z$ defines the
orientation canonical partition function associated with the fundamental 
vibrational level at temperature $T$, $\beta=(k_{B}T)^{-1}$ where $k_B$ is the 
Boltzmann's constant, $\delta$ is the Dirac's function, and $\omega
_{0JMK\rightarrow 1J^{\prime }M^{\prime }K^{\prime }}=\hbar
^{-1}\left( E_{1J^{\prime }M^{\prime }K^{\prime }}-E_{0JMK}\right)$ 
is the transition frequency. Note that we use below the usual spectroscopic 
nomenclature to designate the line transitions connected to the rotational 
selection rules, that is, Q($J_K$) for $\Delta J = 0$, R($J_K$) for $\Delta J = +1$, 
and P($J_K$) for $\Delta J = -1$. 

\subsection{$\nu_2$ mode results and discussion}

In Figure 5, are given the near infrared bar-spectra of NH$_3$ trapped in fullerene 
in the $\nu_2$ mode absorption region for three different temperatures, 10 K, 30 K 
and 100 K. The spectra are calculated by taking into account the frequency shift and 
allowed rovibrational transition elements following selection rules.

For the latter, as the first derivative of the molecular dipole moment with respect 
to the $Q_2$ coordinate does not depend on the rotational spinning angle $\chi$, 
$\Delta K= 0$ for the $K$ quantum number.

The infrared bar-spectrum consists of a Q line located at the pure 
vibration-inversion frequency 751.6 cm$^{-1}$, and a set of 
rotation-vibration-inversion lines P and R 14.8 cm$^{-1}$ apart, corresponding 
to 2$B^{\text{eff}}$, on both sides of the Q branch.

\begin{figure}[h!]
\begin{center}
\includegraphics[width=15cm]{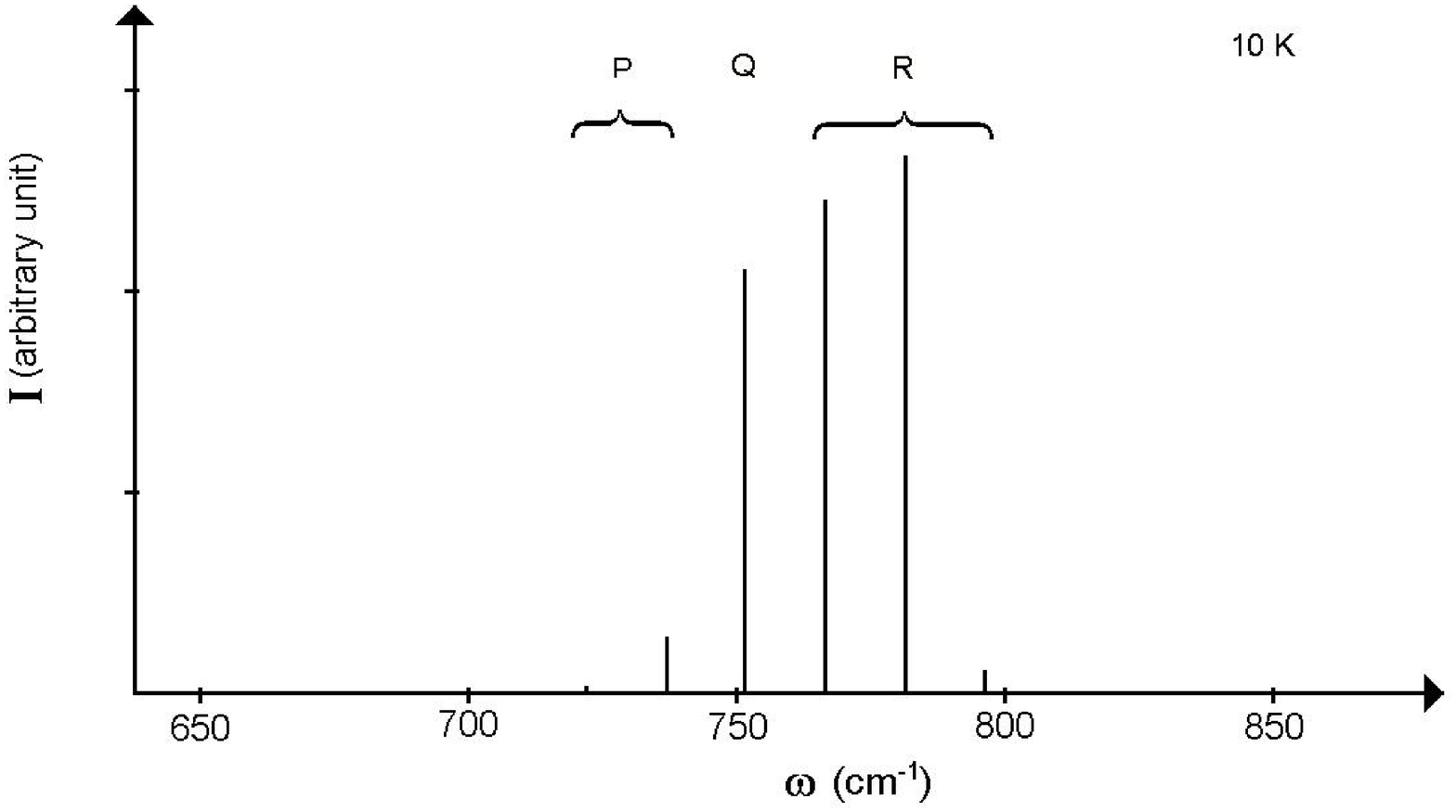}
\end{center}
\begin{center}
\includegraphics[width=15cm]{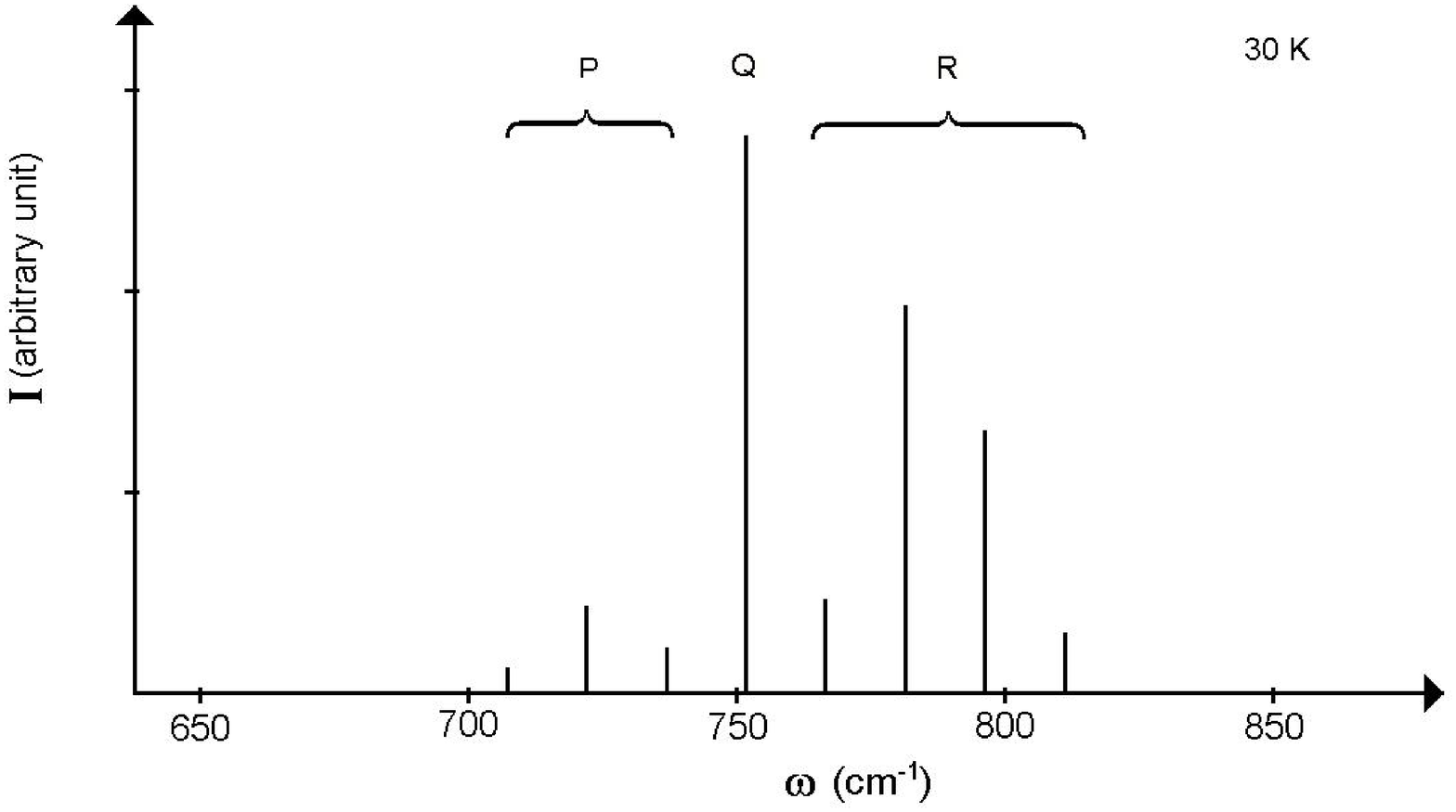}
\end{center}
\begin{center}
\includegraphics[width=15cm]{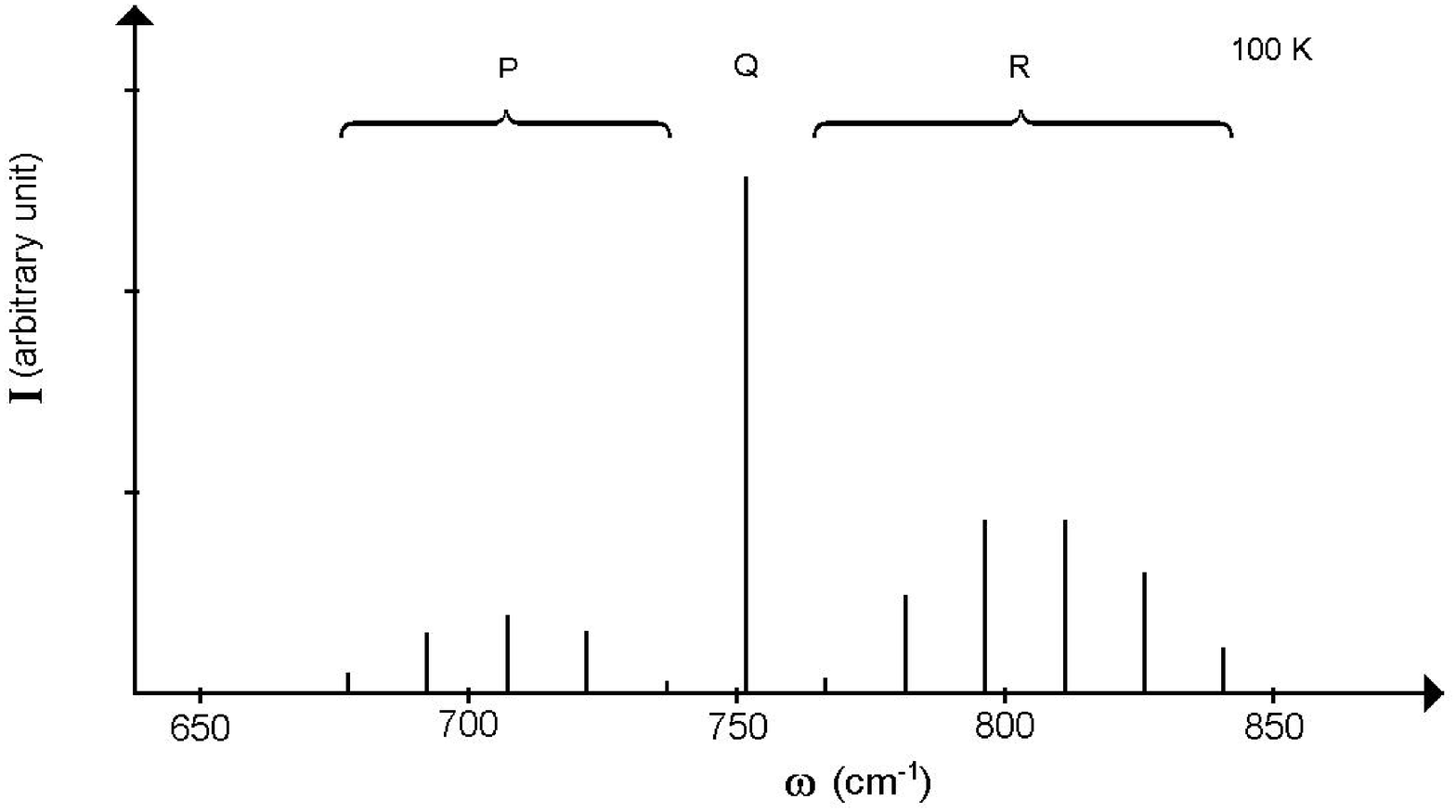}
\end{center}
\caption{Near infrared bar-spectrum in the $\nu_2$ vibration-inversion frequency region for 
NH$_3$ trapped in the fullerene nano-cage at temperatures 10, 30 and 100 K.} 
\label{FIG. 5}
\end{figure}

At $T$ = 10 K, only the lowest few rotational states are populated. As a 
result, the bar-spectrum shown in Figure 5a exhibits only three lines: 
\textit{i}) a Q line at 751.6 cm$^{-1}$ which is in fact a superposition of 
the three transitions Q($1_1$), Q($2_1$) and Q($2_2$), 
\textit{ii}) a R($0_0$) line at 766.4 cm$^{-1}$ 
and \textit{iii}) a R line, superposition of R($1_0$) and R($1_1$) lines 
at 781.3 cm$^{-1}$. There is only one P($1_0$) line 
located at 736.8 cm$^{-1}$.

As the temperature increases, more and more rotational states 
are populated and as result, more lines appear in the spectra with a different 
distribution in terms of intensities. At 30 K, the Q line is the most intense 
consisting of a superposition of Q($J_K$) lines for $J$ = 1 to 4 
and $1\leq K \leq J$. Moreover, the intensity of lines R($0_0$) 
and R($1_K$) decreases leading to the increase of the R($2_K$) line which is located 
at 796.1 cm$^{-1}$. More lines appear in the R and P branches 
(R($3_K$) and P($2_K$)). 

At 100 K, the Q branch consists of a superposition of 6 lines for $J$ varying 
from 1 to 6. The R and P branches are broader with lines of weaker intensities 
than those at 10 and 30 K in particular for the maxima.

The modification of the spectra near $\nu_2$ with temperature shows that the spectral 
signature of the trapped molecule can be used to sense the environmental 
temperature of fullerene C$_{60}$. As the coupling with the vibrational modes 
of C$_{60}$ is weak, the signal to noise ratio may be high enough for all the 
lines to be observed. In our opinion this result can be extrapolated to other 
types of simple molecules trapped in C$_{60}$.

Finally, the dynamical coupling of the vibration-rotation states of the NH$_3$ 
molecule with the vibrational modes of fullerene C$_{60}$ involves well separated 
states for the latter. As a result, this coupling is weak and the energy from 
local translational modes cannot redistribute itself in the thermal bath of C$_{60}$.
The linewidths are then negligible. 

In conclusion, this work shows that the simulation of the spectra of NH$_3$ can 
be used to probe the temperature of the media in which fullerene is observed. 
This result can probably be extended to other molecules trapped in fullerene, in 
particular when the coupling with the vibration modes of fullerene is weak, because 
signal to noise ratio should then be high enough for the rotation lines to be observed. 

\newpage

\textbf{ACKNOWLEDGEMENTS}

The authors thank the Regional Councels of Franche-Comt\'{e} and Ile de France, 
the General Councel of the 78$^{\text{th}}$ district of Yvelines, the DGCIS 
from Ministry of Industry in France, the MOVEO cluster for financing the FUI 
project MEMOIRE under the label of cluster MOVEO, CNRS for financing support under 
the Interdisciplinary Research Program of the INSU \textquotedblleft Planetary 
Environments and Life Origins (EPOV)\textquotedblright, and also Camille Lakhlifi 
for figures help.

\appendix

\section{Kinetic Lagrangian expression}

The kinetic Lagrangian associated with the ammonia inversion, orientation, and 
c.m. translation modes considered as classical motions is 

\begin{equation}
\textit{L}_{\text{K}} = \frac{1}{2}\mu_2 \dot{Q_2}^2 + \frac{1}{2}I_{\text{B}}%
(\dot{\theta}^2 + \dot{\varphi}^2 \sin^{2} \theta) + \frac{1}{2}I_{\text{C}}%
(\dot{\varphi} \cos \theta + \dot{\chi})^2 + \frac{1}{2}M \dot{r}_{0}^{2}, \label{eqA1}
\end{equation}

\noindent where $\mu_2$ is the reduced mass of the molecule associated with the 
$\nu_2$ mode, $I_{\text{B}}$ and $I_{\text{C}}$ are the molecular principal moments 
of inertia for the rotation motion of its C$_3$ symmetry axis and the spinning motion around this 
axis, respectively, and \textit{M} is the molecular mass. 
Moreover, as it has been mentioned above, the position vector 
of the centre of mass of the molecule can be written as 
\textbf{r}$_{0}$ = \textbf{r}$_{0}^{\text{e}}(\varphi,\theta,Q_2)$ + \textbf{u}$_{0}$. 
The velocity vector $\dot{\textbf{r}}_{0}$ is then 

\begin{equation}
\dot{\textbf{r}}_{0} = (\dfrac{\partial\textbf{r}_{0}^{\text{e}}}{\partial \varphi} \dot{\varphi} + 
\dfrac{\partial\textbf{r}_{0}^{\text{e}}}{\partial \theta} \dot{\theta} +  
\dfrac{\partial\textbf{r}_{0}^{\text{e}}}{\partial Q_2}\dot{Q_2}) +
\dot{\textbf{u}}_{0}. \label{eqA2}
\end{equation}

By substituting Eq. (\ref{eqA2}) into Eq. (\ref{eqA1}), this latter equation becomes

\begin{equation}
\textit{L}_{\text{K}} = \frac{1}{2}\mu_{2}^{\text{eff}} \dot{Q_2}^2 + \frac{1}{2}I_{\text{B}}%
^{\text{eff}}(\dot{\theta}^2 + \dot{\varphi}^2 \sin^{2} \theta) + \frac{1}{2}I_{\text{C}}%
(\dot{\varphi} \cos \theta + \dot{\chi})^2 + \frac{1}{2}M \dot{u}_{0}^{2} + 
M\dfrac{\partial r_{0}^{\text{e}}}{\partial Q_2}\dot{Q_2} \dot{u}_{0}, \label{eqA3}
\end{equation}

\noindent where $\mu_{2}^{\text{eff}}$ {and} $I_{\text{B}}^{\text{eff}}$ are effective 
reduced mass and moment of inertia 

$\mu_{2}^{\text{eff}} = \mu_2 + M \left(\dfrac{\partial r_{0}^{\text{e}}}
{\partial Q_2}\right)^2$,  \text{    and    }   $I_{\text{B}}^{\text{eff}} = I_{\text{B}} + 
M {r_{0}^{\text{e}}}^{2}$.

Note that the last term in Eq. (\ref{eqA3}) accounts for the simultaneous dynamical 
evolution of the translation motion of the centre of mass of the molecule and 
its inversion motion. 

\section{Rotational matrix transformation and molecular dipole moment}

The unitary matrix $\textbf {M}$ characterizing the transformation from
the absolute frame (O,\textbf{X},\textbf{Y},\textbf{Z}) into the molecular one 
(G,\textbf{x},\textbf{y},\textbf{z}) through the Euler 
angles $\varphi $, $\theta $ and $ \chi $ is given as \cite{mer67}\\

$$
\mathbf{M}(\varphi ,\theta ,\chi )=
$$

\begin{equation}
\left(
\begin{array}{ccc}
\cos \varphi \cos \theta \cos \chi -\sin \varphi \sin \chi & \sin \varphi
\cos \theta \cos \chi +\cos \varphi \sin \chi & -\sin \theta \cos \chi \\
-\cos \varphi \cos \theta \sin \chi -\sin \varphi \cos \chi & -\sin \varphi
\cos \theta \sin \chi +\cos \varphi \cos \chi & \sin \theta \sin \chi \\
\cos \varphi \sin \theta & \sin \varphi \sin \theta & \cos \theta
\end{array}
\right) . \label{eqB1}
\end{equation}

Thus, the molecular dipole moment $\mu _A$ and its first derivative vectors 
$\frac{\partial \mu _A}{\partial Q_\nu }$, with respect to the normal coordinate 
$Q_\nu $ of the $\nu $th vibrational mode can be written in the absolute
frame as

\begin{eqnarray}
\mathbf{\mu }_A &=&\mathbf{M}^{-1}(\varphi ,\theta ,\chi )\mathbf{\mu },
\nonumber \\
\frac{\partial \mathbf{\mu }_A}{\partial Q_\nu } &=&\mathbf{M}^{-1}(\varphi
,\theta ,\chi )\mathbf{b}^\nu , \label{eqB2}
\end{eqnarray}

\noindent where $\textbf {M}^{-1}$ is the inverse matrix of 
$\textbf {M}$, and $\textbf {b}^\nu$ = $\frac{\partial {\mu }}{\partial Q_\nu }$ 
the first derivative vectors of the molecular dipole moment expressed in the 
molecular frame. Their non vanishing component values (in D$\text{\AA}^{-1}$) 
are \cite{aljpk2005}

\begin{eqnarray}
\left\{
\begin{tabular}{lr}
$\text{b}^1=\frac{\partial \mu _z}{\partial Q_1}=1.404$ & $\text{b}^2=\frac{\partial \mu _z%
}{\partial Q_2}=-4.409$ \\
$\text{b}^{3a}=\frac{\partial \mu _x}{\partial Q_{3a}}=-4.012$ & $\text{b}^{3b}=\frac{%
\partial \mu _y}{\partial Q_{3b}}=2.465$ \\
$\text{b}^{4a}=\frac{\partial \mu _x}{\partial Q_{4a}}=1.984$ & $\text{b}^{4b}=\frac{%
\partial \mu _y}{\partial Q_{4b}}=-2.559$
\end{tabular}
\right\}. \nonumber
\end{eqnarray}

Moreover, the values of the elements of the normal coordinates connected to the 
fundamental vibrational transitions are (in $\text{\AA}$)

\begin{eqnarray}
\left\{
\begin{tabular}{l}
$\left| \left\langle 0_1 \left| Q_1 \right|
1_1\right\rangle \right|=0.042$ \\
$\left| \left\langle 0_2 \left| Q_2 \right|
1_2\right\rangle \right|=0.069$ \\
$\left| \left\langle 0_3 \left| Q_3 \right|
1_3\right\rangle \right|=0.029$ \\
$\left| \left\langle 0_4 \left| Q_4 \right|
1_4\right\rangle \right|=0.060$
\end{tabular}
\right\}. \nonumber
\end{eqnarray}

\end{document}